\documentclass[11pt,twoside]{article}
\usepackage{asp2010}

\resetcounters

\aspvolume{455}
\aspcpryear{2012}
\aspvoltitle{4$^{th}$ {\em Hinode} Science Meeting: Unsolved 
Problems and Recent Insights}
\aspvolauthor{L.~R.~Bellot Rubio, F.~Reale, and M.~Carlsson, eds.}

\begin{document}

\title{Propagating Intensity Disturbances in Fan-like Coronal Loops: Flows or Waves?}
\author{Tongjiang~Wang,$^{1,2}$ Leon~Ofman,$^{1,2,3}$ and Joseph~M.~Davila$^2$
\affil{$^1$Department of Physics, Catholic University of America, 620 Michigan Avenue, Washington, DC 20064, USA}
\affil{$^2$NASA Goddard Space Flight Center, Code 671, Greenbelt, MD 20771, USA}
\affil{$^3$Visiting Associate Professor, Tel Aviv University, Israel}
}

\begin{abstract}
Quasi-periodic intensity disturbances propagating upward along the
coronal structure have been extensively studied using EUV imaging
observations from SOHO/EIT and TRACE. They were interpreted as either
slow mode magnetoacoustic waves or intermittent upflows. In this study
we aim at demonstrating that time series of spectroscopic observations
are critical to solve this puzzle. Propagating intensity and Doppler
shift disturbances in fanlike coronal loops are analyzed in multiple
wavelengths using sit-and-stare observations from {\em Hinode}/EIS.  We
find that the disturbances did not cause the blue-wing asymmetry of
spectral profiles in the warm ($\sim$1.5~MK) coronal lines.  The
estimated small line-of-sight velocities also did not support the
intermittent upflow interpretation. In the hot ($\sim$2~MK) coronal
lines the disturbances did cause the blue-wing asymmetry, but the
double fits revealed that a high-velocity minor component is steady
and persistent, while the propagating intensity and Doppler shift
disturbances are mainly due to variations of the core component,
therefore, supporting the slow wave interpretation.  However, the
cause for blueward line asymmetries remains unclear.

\end{abstract}

\section{Introduction}
Upwardly propagating quasi-periodic intensity disturbances are known
to exist in coronal loops \citep{1999SoPh..186..207B,
2000A&A...355L..23D} and have been studied extensively using EUV
images from SOHO/EIT and TRACE \citep[see a recent review
by][]{2009SSRv..149...65D}.  These propagating disturbances were
interpreted as slow magnetoacoustic waves
\citep{2000A&A...362.1151N} since their propagating speeds 
($\sim$100~km~s$^{-1}$) are close to the sound speed in the corona and
periodicities appear depending on the underlying driver such as the
leakage of the photospheric $p$-modes \citep{2002SoPh..209...61D,
2005ApJ...624L..61D}.

\begin{figure}[!ht]
\plotone{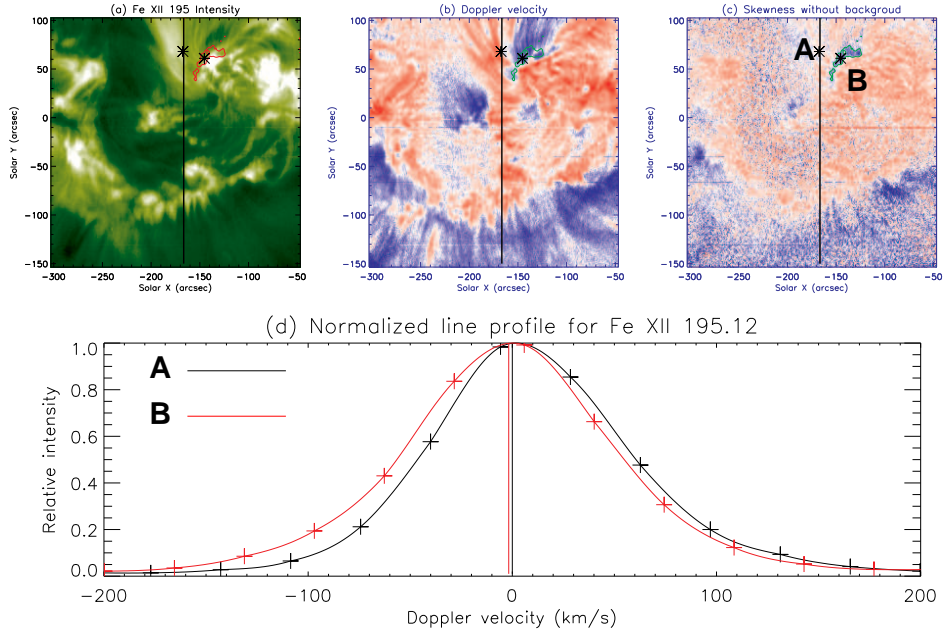}
\caption{{\em (a)} EIS \ion{Fe}{xii}~195~\AA\ raster image. 
{\em (b)} Doppler velocity map.  {\em (c)} Skewness map with a
constant background subtracted.  The contours shown in (a)--(c)
indicate an area with the large negative skewness, and the thick
vertical line marks the slit position for the sit-and-stare
observations.  {\em (d)} The normalized line profiles at positions A
(in black) and B (in red) marked in (c).  \label{fgfe12}}
\end{figure}

However, similar phenomena were re-discovered with {\em Hinode}/XRT
recently and interpreted as high-speed outflows because these
disturbances were found associated with outflows at the loop's
footpoints with {\em Hinode}/EIS \citep{2007Sci...318.1585S,
2008ApJ...676L.147H}.  Based on EIS observations of blueward
asymmetries of spectral profiles, some authors suggested that these
propagating disturbances are most likely transient outflows produced
by episodic coronal heating events \citep{2009ApJ...706L..80M,
2009ApJ...701L...1D, 2010ApJ...722.1013D}.  However, the blueward
spectral asymmetry is also consistent with the wave interpretation
\citep{2010ApJ...724L.194V}.

\begin{figure}[!ht]
\plotone{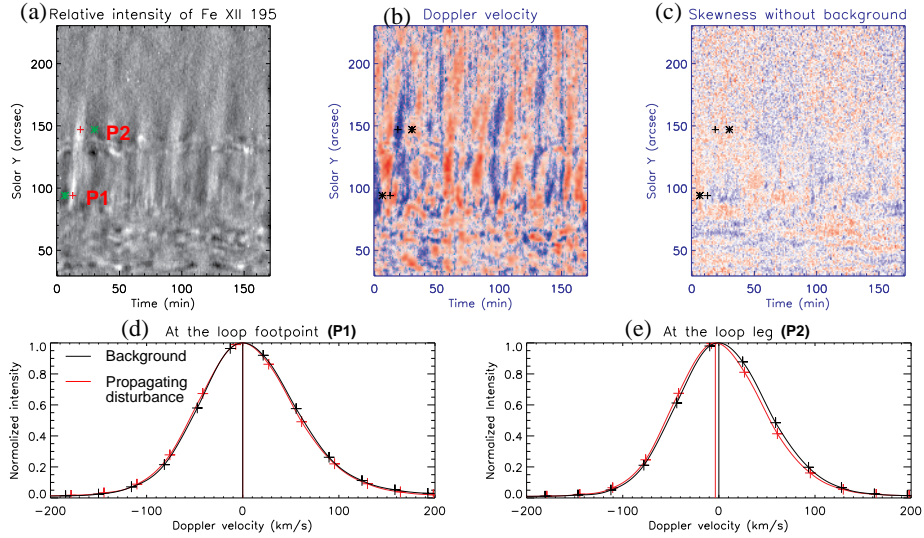}
\caption{Propagating disturbances in coronal loops observed with 
EIS in the Fe\,{\sc{xii}} 195\AA\ line.  {\em (a)} Time series of
relative intensity along the slit with a scale ranging from $-$10\%
(dark) to $+10\%$ (bright).  {\em (b)} Time series of relative Doppler
velocity with a scale ranging from $-3$ to $+3$~km~s$^{-1}$.  {\em
(c)} Time series of skewness with removal of a background which is
taken as an average over time at each slit position. The scale range
is from $-0.3$ to $+0.3$. {\em (d)} and {\em (e)} Comparisons of the
normalized line profiles at positions P1 and P2. The red line
represents the spectral line profile at the bright disturbance, while
the black line shows the profile in between (referred to as the
background).
\label{fgpdfe12}}
\end{figure}

\citet{2009A&A...503L..25W} first studied
these propagating disturbances using the EIS time series and found
that the Doppler shift variations have small amplitudes of only about
1$-$2 km~s$^{-1}$, and are in phase with the intensity
variations. These features are consistent with slow magnetoacoustic
waves propagating upwards along coronal structures.  In this study we
examine the effect of the disturbances on the evolution of spectral
profiles.  The purpose is to determine whether the apparent
propagating disturbances are the slow waves or high-velocity upflows,
or possible coexistence of both.

\section{Observations and Data Analysis}
The observations were obtained on February 1, 2007 in AR 10940.  The
analysis of EIS sit-and-stare observations for propagating
disturbances in Fe\,{\sc{xii}} 195.12 \AA\ was presented by
\citet{2009A&A...503L..25W}. They found quasi-periodic (12 and 25
min) propagating disturbances in both intensity and Doppler shift
along the fanlike loops.
\citet{2010ApJ...722.1013D} performed a ``Red Minus Blue'' 
(R-B) profile asymmetry analysis on the Fe\,{\sc{xii}} 195.12,
Fe\,{\sc{xiii}} 202.04, and Fe\,{\sc{xiv}} 274.20 \AA\/ spectral
lines, and discussed the cause of the quasi-periodicities in the R-B
asymmetry using the forward modeling and R-B guided double fits
techniques.  Using a different approach, our study focuses on direct
examination of the influence of individual disturbances on the line
profile shape. We also examine the line asymmetry using the skewness.
The skewness is a typical measure of line asymmetry, defined as
$\int[(\lambda-\lambda_0)/\sigma]^3I(\lambda){\rm
d}\lambda/\int{I}(\lambda){\rm d}\lambda$, where $\lambda_0$ and
$\sigma$ are the mean and standard deviation of the line.
Figure~\ref{fgfe12} shows the Fe\,{\sc{xii}} raster image, Doppler
shift, and skewness maps. Since the Fe\,{\sc{xii}} 195.12 \AA\ line is
self-blended with a weak line \citep[195.18
\AA;][]{2009A&A...495..587Y}, a constant background has been removed
from the skewness map.  The skewness with large negative values
indicates line profiles having evident blueward asymmetries
(Figures~\ref{fgfe12}c and d). To find the change of a spectral
profile in shape relative to the background profile, we first
interpolate the spectral profile, and then normalize it to the peak
intensity. We calibrate Doppler velocities measured with EIS in
different emission lines by choosing Si\,{\sc{x}} 258.37 \AA\ in
the long wavelength band and Fe\,{\sc{xii}} 195.12 \AA\ in the
short wavelength band as the reference lines and adopting the relative
wavelengths measured above the limb by \citet{2011ApJ...727...58W}.
We adjust the wavelength scale of the reference lines by letting the
quiet corona have no net velocities in these lines.

\begin{figure}[!ht]
\plotone{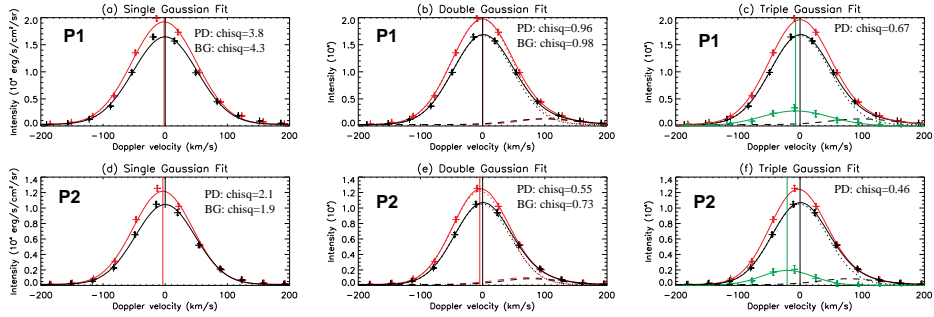}
\caption{Comparison of the line profiles of Fe\,{\sc{xii}} 195.12 \AA\ 
on and off a bright propagating disturbance (PD) at position P1 ({\em
top}) and P2 ({\em bottom}) along the slit shown in
Figure~\ref{fgpdfe12}a.  {\em (a)} Single Gaussian fit to the line
profiles (red for PD, black for the background).  {\em (b)} Same as
(a), but for double Gaussian fit with the second component for a
self-blended line, Fe\,{\sc{xii}} 195.18 \AA. {\em (c)} Triple
Gaussian fit to the PD line profile (red line), by taking the
background profile (fitted with the double Gaussian in solid black
line) as a fixed component, where the excess component (in solid green
line) is expected from the {\em hypothetical} upflow.  {\em (d)--(f)}
Same as (a)--(c), but for line profiles at position P2. In each
panel, the reduced chi-squared statistics of the fits are
marked. \label{fgftfe12}}
\end{figure}

\section{Results for Spectral Profile Examination in Multiple Wavelengths}

\subsection{Warm Coronal Lines: Fe~\textsc{xii} 195 \AA\/ and 
Fe~\textsc{xiii} 202 \AA}

Figure~\ref{fgfe12} shows that the strong blueward asymmetry of
Fe\,{\sc{xii}} 195 \AA\ mainly occurred at the footpoints of weak
loops associated with upflows, while in bright fanlike loops
associated with downflows, signatures of the blueward asymmetry are
less evident.  Figure~\ref{fgpdfe12} shows the propagating
disturbances in intensity and Doppler shift of Fe\,{\sc{xii}} along
the bright fanlike loops. No propagating features found in time series
of the skewness indicate that the disturbances did not cause changes
in the line shape.  This is confirmed by comparisons of the normalized
line profiles at times on and off the propagating brightenings
(Figures~\ref{fgpdfe12}d and \ref{fgpdfe12}e).

The line asymmetry also can be examined by the multiple fit
analysis. Figure~\ref{fgftfe12} shows that single Gaussian fits did
not fully capture the data (with large reduced chi-squared statistics)
because of the blending. The double Gaussian fits including the
blended line of Fe\,{\sc{xii}} 195.18 \AA\ match well to the data. For
such line profiles, it is impossible to add an extra component in free
multiple Gaussian fits. To check the suggestion by
\citet{2010ApJ...722.1013D} that bright disturbances are caused by the
quasi-periodic transient upflows, we make the {\em hypothesis} that
the excess of the line profile for a bright disturbance (also called
transient jets by some authors) relative to that taken before or after
this disturbance at the same position (referred to as the background)
is fully contributed by an upflow, and then we can make a multiple
Gaussian fit to the line profile for the disturbance by fixing the
background components (Figures~\ref{fgftfe12}c and
\ref{fgftfe12}f). From the fits, we can estimate the line-of-sight
velocity of the {\it hypothetical} upflow, which is found to be about
7 and 20 km s$^{-1}$ relative to the background component for the
positions at the footpoint and the leg of the loops, respectively.  We
also notice that the Gaussian width of the fitted upflow component is
larger than that of the background component by 16\% at the position
of the loop's footpoint, whereas it is smaller than that of the
background component by 8\% at the position of the leg.

\begin{figure}[!ht]
\plotone{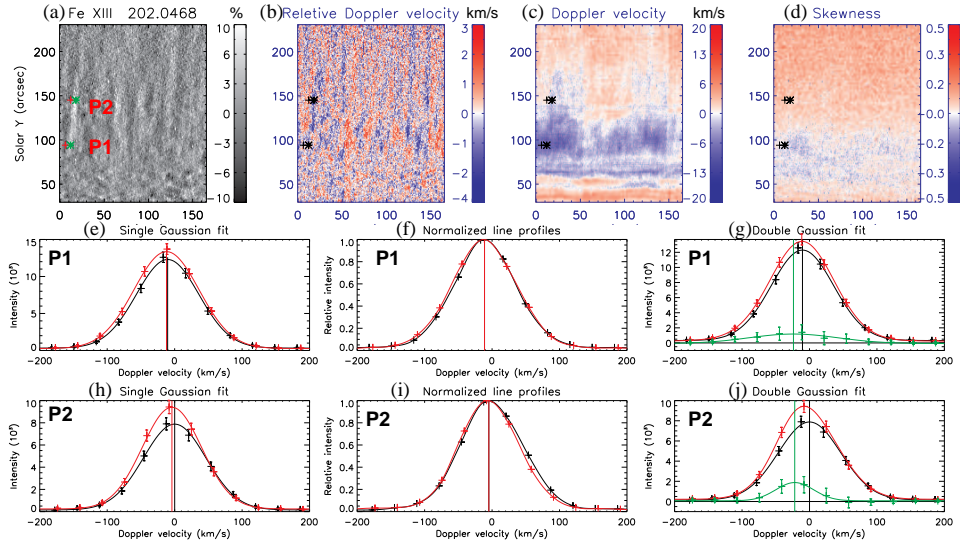}
\caption{Time series of {\em (a)} relative intensity, {\em (b)} 
relative Doppler shift with the background trend subtracted, {\em (c)}
Doppler shift, and {\em (d)} skewness in Fe\,{\sc{xiii}} 202.05 \AA\
for the propagating disturbances in coronal loops observed with
EIS. {\em (e)} Line profiles along with single Gaussian fits at P1
(red line for PD, black line for the background).  {\em (f)}
Comparison between the normalized line profiles on and off the PD.
{\em (g)} The PD line profile along with Double Gaussian fit (red
line), by taking the background line profile (fitted with single
Gaussian in black line) as a fixed component, where the excess
component (green line) is expected from an {\em hypothetical}
upflow. {\em (h)-(j)} Same as (e)-(g), but for position P2.
\label{fgftfe13}}
\end{figure}

Although the Fe\,{\sc{xii}} 195 \AA\ observations have the highest
signal-to-noise ratio, the self-blending made its analysis
complicated as shown above. To confirm the results found in
Fe\,{\sc{xii}}~195 \AA, we compare with the analysis of a clean line,
Fe\,{\sc{xiii}}~202.05 \AA. Figures~\ref{fgftfe13}a and
\ref{fgftfe13}b show that the propagating intensity and Doppler shift 
disturbances observed in Fe\,{\sc{xiii}} are very close to those seen
in Fe\,{\sc{xii}}.  The negative skewness in the region $y =
60\arcsec$--$120\arcsec$ indicates that the blueward line asymmetry
may result from the associated upflows in this region
(Figures~\ref{fgftfe13}c and \ref{fgftfe13}d).  However, the
disturbances can propagate high above the region of the blueward
asymmetry, and the skewness evolution shows no evidence for similar
propagating features. These facts suggest that the upflows near the
footpoints of the loop are persistent and stable during the
observations, which cannot account for the quasi-periodic disturbances
propagating to a high level. The line profile analysis shows that the
estimated line-of-sight velocity of the {\it hypothetical} upflow is
about 13 km s$^{-1}$ at the footpoint (Fig.~\ref{fgftfe13}g), while
about 22 km s$^{-1}$ at the leg of the loop (Figure~\ref{fgftfe13}j).
This result agrees well with that for Fe\,{\sc{xii}}. The line width
of the flow component estimated for Fe\,{\sc{xiii}} is 42\% larger at
the footpoint, but 40\% smaller at the leg of the loop than that of
the background, which follows the same rule as for Fe\,{\sc{xii}}. In
addition, the line-of-sight velocity of the {\it hypothetical} upflows
estimated for Si\,{\sc{x}} 258.37\AA\ (1.3 MK) are 11--13~km~s$^{-1}$,
also consistent with the above results.

Under the hypothesis that the propagating disturbances are transient
upflows, the line-of-sight component of 10--20~km~s$^{-1}$ along with
the transverse component (assumed to be the observed propagating
speed) of 100 km~s$^{-1}$ implies an inclination of the magnetic field
in coronal loops to the vertical in the range 79\deg--84\deg, which
is too large to be acceptable. \citet{2009ApJ...697.1674M} measured
the 3D geometry of fanlike loops using STEREO EUVI observations, and
found their inclination to be $37\deg \pm 6\deg$ to the local
normal. Moreover, the fact that the line width of the {\it
hypothetical} upflow at the footpoint of the loop is much larger than
that at the leg is unexpected, while this may be explained by the
presence of multiple harmonics of the waves
\citep{2009A&A...503L..25W}. Therefore, the above upflow hypothesis is not supported, instead the wave interpretation is
preferred.  

\begin{figure}[!t]
\plotone{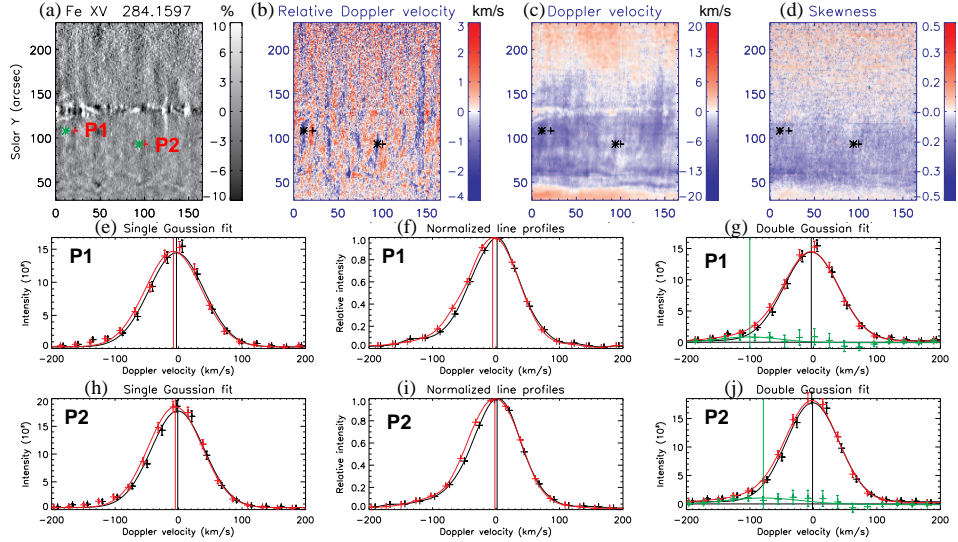}
\caption{Same as Fig.~\ref{fgftfe13}, but for the 
Fe\,{\sc{xv}} 284.16 \AA\ line. \label{fgftfe15}}
\end{figure}

\subsection{Hot Coronal lines: Fe~\textsc{xiv}~274 \AA\/ and 
Fe~\textsc{xv}~284 \AA}

The propagating disturbances along hot coronal loops in the
Fe\,{\sc{xiv}} 274.20 \AA\ and Fe\,{\sc{xv}} 284.16 \AA\ emission
lines look similar, but appear different from those seen in the warm
lines discussed in the last section.  We find that the propagating
disturbances detected in intensity and Doppler shift between warm and
hot coronal lines are less correlated (not shown).
Figures~\ref{fgftfe15}a and
\ref{fgftfe15}b show that the propagating disturbances are weak in
intensity, but more evident in Doppler shift and located near the
footpoints of the loop. The presence of persistent upflows is revealed
in these hot coronal lines, which are characterized by evident
blueward line asymmetries in Fe\,{\sc{xv}} indicated by the skewness
(Figures~\ref{fgftfe15}c and \ref{fgftfe15}d). Single Gaussian fits show
evident excess emission in the blue wing above 100~km~s$^{-1}$
(Figures~\ref{fgftfe15}e and \ref{fgftfe15}h).  Comparisons of the
normalized line profiles reveal the enhancement in the blue wing of
Fe\,{\sc{xv}} due to the disturbances (Figures~\ref{fgftfe15}f and
\ref{fgftfe15}i). But this feature was not found in the warm coronal
lines (Figures~\ref{fgftfe13}f and \ref{fgftfe15}i).

\begin{figure}[!ht]
\plotone{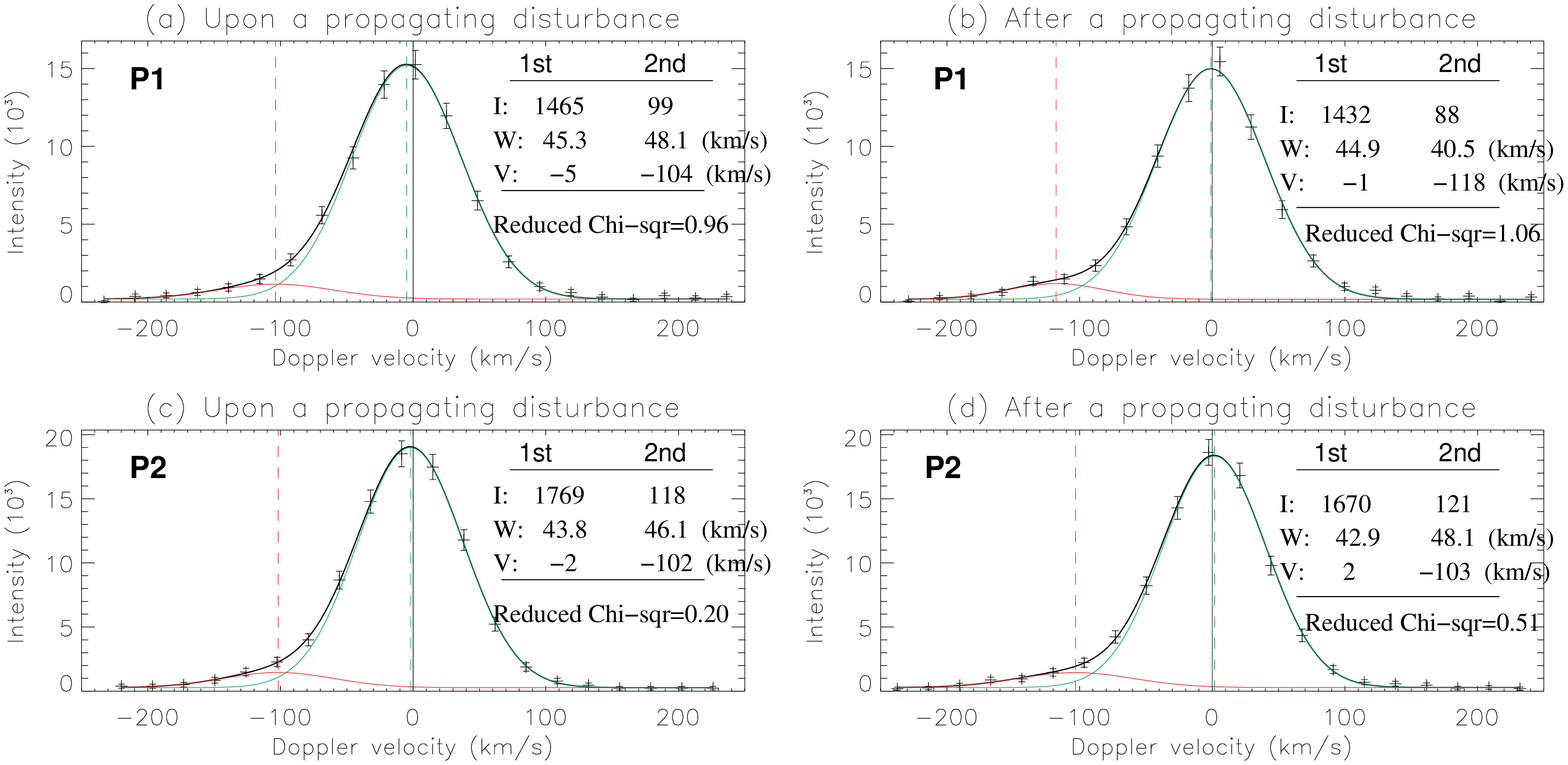}
\caption{Spectral profiles of Fe\,{\sc{xv}} 284.16 \AA\ recorded 
in one spatial pixel at position P1 along the slit {\em (a)} upon and
{\em (b)} after a propagating disturbance as shown in
Figure~\ref{fgftfe15}. {\em (c)--(d)} Same as (a)--(b), but for
position P2.  The constrained double Gaussian fits were made by
forcing the minor component (red line) to have a blueshift of at least
70 km s$^{-1}$. Vertical dashed lines indicate the peak positions of
the two fitted components. The measured intensity (I), line width (W),
and Doppler shift (V) for the fits are marked in each
panel. \label{fg2gft}}
\end{figure}

Using the double fits by taking the background line profile as a fixed
component, we estimate the line-of-sight velocity of the {\it
hypothetical} upflow to be 97 and 78 km~s$^{-1}$ for two selected
disturbances (Figures~\ref{fgftfe15}g and \ref{fgftfe15}j).  However,
it is noticed that the line profiles for both the disturbance and the
background have evident blueward line asymmetries, indicating the
presence of a persistent upflow. We use the constrained double
Gaussian fits as suggested in \citet{2010A&A...521A..51P} to analyze
the effect of the disturbances on the evolution of the core component
and the second minor component.  Figure~\ref{fg2gft} shows the line
profiles of Fe\,{\sc{xv}} 284 \AA\ along with the double fits upon and
after the disturbances, where the second minor component was forced to
have a blueshift of at least 70 km~s$^{-1}$. We find that the
intensity increase of the core component accounts for about 75\% of
the total increase in Fe\,{\sc{xv}} intensity for case P1, while it 
is more than the total increase for case P2.  The core component was
blueshifted by 4 km~s$^{-1}$ due to the bright disturbance in both
cases, while the Doppler velocity of the minor component kept at about
100$-$120 km~s$^{-1}$ with a small decrease upon the disturbance
relative to the background (taken after the disturbances). This result
indicates that the propagating intensity and Doppler shift
disturbances with small amplitudes obtained in single fits are mainly
ascribed to variations of the core component (loop emission) but not
of the second component (high-speed upflows) in double fits for the
hot coronal lines.  Therefore, our result supports the interpretation
of the propagating disturbances as slow mode magnetoacoustic waves
propagating in coronal loops.

\section{Discussion and Conclusion}
We have presented a detailed analysis of propagating intensity and
Doppler shift disturbances in fanlike coronal loops using EIS
sit-and-stare multi-wavelength observations. We adopted several
techniques such as the skewness, the line profile normalization, and
the multiple Gaussian fits to examine changes of the spectral profile
shape caused by these disturbances in warm coronal lines (e.g.,
Fe\,{\sc{xii}} 195 \AA\ and Fe\,{\sc{xiii}} 202
\AA) and in hot coronal lines (e.g., Fe\,{\sc{xv}} 284 \AA).

Blue-wing asymmetries associated with Doppler blueshifts are found to
exist near the footpoints of coronal loops from the skewness analysis,
they are more evident in hot lines. No propagating features found in
time series of the skewness suggest that the associated upflows are
persistent during the observations.  We find that these disturbances
appear uncorrelated between the warm and hot lines, suggesting that
they may propagate in different loops of different temperatures.  The
warm coronal lines show no evidence that the disturbances led to
blue-wing asymmetries.  The observed high (100--120~km~s$^{-1}$)
propagating speeds are not likely to be the result of hypothetical
periodic upflows whose line-of-sight velocity is estimated to be only
about 10--20~km~s$^{-1}$. The hot coronal lines have a high-velocity
($\sim$100~km~s$^{-1}$) component persistently existing in the blue
wing during the period of observations, suggesting the presence of a
steady upflow in hot coronal loops. The restricted double fits show
that the propagating intensity enhancements are mainly contributed
from the increase in the core component but not due to variations of
the second minor component. Variations of the core component in
intensity and Doppler shift having small amplitudes are consistent
with propagating slow-mode waves. However, the disturbance-related
excess in the blue wing of hot coronal lines revealed by comparisons
of the normalized line profiles (Figures~\ref{fgftfe15}f and
\ref{fgftfe15}i) could not be explained by the superposition of the
core component with Doppler shift oscillations caused by a slow wave
and the background component produced by a steady upflow or static
plasma based on a simple forward modeling, and the cause requires
further investigation.

Long-period (12 and 24 min) slow waves are not likely to be the direct
result of photospheric $p$-mode leakage. Instead their association
with steady upflows may suggest that these waves are produced by the
same mechanism that drives the steady upflows.  As these upflows are
progressively larger in lines formed at higher temperatures,
\citet{2008A&A...481L..49D} suggested that they could be due to 
chromospheric evaporation following magnetic reconnection. The
generation of slow waves is expected since they are a natural response
to the impulsive heating \citep{2004A&A...414L..25N,
2007ApJ...659L.173T}. Therefore, observations of these waves may
provide a potential diagnostic of coronal heating. More recently,
\citet{2011A&A...526A.137D} proposed a rarefaction wave model based on
reconnection between field lines of active regions and the nearby
coronal hole, which can also explain the existence of persistent
upflows and slow waves. But the reason why the upflows are persistent
while the waves appear as quasi-periodic pulses is unknown and should
be further studied in the future.

\acknowledgements 
{\em Hinode} is a Japanese mission developed, launched, and operated
by ISAS/JAXA in partnership with NAOJ, NASA, and STFC (UK). Additional
operation support is provided by ESA and NSC (Norway). The work of TW
and LO was supported by NASA grants NNX10AN10G and NNX09AG10G.

\bibliography{hinode4}
\end{document}